\documentclass[12pt]{article}
\usepackage{graphicx,color,amssymb,amsfonts}
\textheight by \topskip \textwidth 455pt \hfuzz=15pt

\newcommand{\AmS}{{\protect\the\textfont2
A\kern-.1667em\lower.5ex\hbox{M}\kern-.125emS}} 
 \normalsize 
\begin{document}
\centerline{\bf EAS Longitudinal Development and the
Knee\footnote{Talk given at 30th ICRC (2007), Merida, Mexico}}
\vspace{12pt} \centerline{\it Yuri V. Stenkin}
\vspace{12pt}{\footnotesize\it
Institute for Nuclear research of Russian Academy of Sciences}\\
{\footnotesize\it 60th October anniv. prospect,
7a, Moscow 117312, Russia}\\
\footnotesize{e-mail: stenkin@sci.lebedev.ru}
\begin{abstract}
It is shown that Extensive Air Shower (EAS) longitudinal
development has a critical point where an equilibrium between the
main hadronic component and the secondary electromagnetic one
exhibits a brake. This results in a change of slope in quasi-power
law function $N_{e}(E_{o})$. The latter leads to a knee in the EAS
size spectrum at primary energy of about 100 TeV/nucleon. Many
``strange'' experimental results can be successfully explained in
the frames of current approach.
\end{abstract}
\section{Introduction}

In 1958 there was claimed $\cite{KK}$ the existence of the
``knee'' in primary cosmic ray spectrum and its possible
explanation by an existence of Galactic and Extragalactic cosmic
rays. It should be noted that in the cited paper there was no any
doubt that visible break in EAS size spectrum could be connected
with any other reason but with primary spectrum steepening. At
those times the recalculation from EAS size to primary energy was
very simple. People merely used a constant coefficient for
recalculation from EAS size at maximum to primary energy. Other
ground level experiments later confirmed the ``knee'' existence in
EAS size spectrum while direct measurements of primary cosmic ray
nuclei spectra at satellites and balloons made up to energy
\ensuremath{\sim}1 PeV do not confirm deviation from a pure power
law at energies above 10 TeV. All experimental data confirming the
``knee'' existence are originated from \textit{indirect
measurements} using the EAS technique. Some physicists tried to
explain the visible knee by a dramatic change in parameters of
particle interactions $\cite{Nik,kaz,pet}$. Absolutely new
approach to this problem has been proposed in 2003 and details of
the approach can be found elsewhere $\cite{st1}$. It has been also
shown $\cite{st2,st3}$ that a lot of experiments contradict the
hypothesis of the ``knee'' in primary spectrum and its
astrophysical origin.

\section{The EAS method}

 An advantage of the EAS method is a possibility to work
up to the highest energy. But, the indirect measurements have to
be recalculated to primary spectrum. This is a very complicated
and model dependent problem. If the primary spectrum follows a
power law function of a type:
I$\ensuremath{\sim}E_{0}^{\ensuremath{-\gamma}}$ and a secondary
component N$_{x}$ also follows a power law:
N$_{x}\ensuremath{\sim} E_{0}^{\ensuremath{\alpha}}$, then
I\ensuremath{\sim}N$_{x}^{\ensuremath{-\beta}}$, where
\ensuremath{\beta=\gamma/\alpha}. If a break in a power law of
experimental data distribution exists , then a change in any of
the two indices (\ensuremath{\gamma} or \ensuremath{\alpha}) may
be responsible for this.

Suppose the primary spectrum index \ensuremath{\gamma} changes at
a point E$_{0}$=E$_{knee}$ from \ensuremath{\gamma} to
\ensuremath{\gamma}+\ensuremath{\Delta\gamma}. Then, one could
expect a predictable break in the index \ensuremath{\beta} for
each component: \ensuremath{\Delta\beta}=\ensuremath{\Delta \gamma
/ \alpha}. Typical values for \ensuremath{\alpha} are the
following: \ensuremath{\alpha}$_{e}$\ensuremath{\approx}1.1-1.25
for electron component and
\ensuremath{\alpha}$_{h}$\ensuremath{\approx}0.8 - 0.9 for
hadronic and muonic components. If \ensuremath{\Delta\gamma}=0.5,
then expected values are:
\ensuremath{\Delta\beta}$_{e}$\ensuremath{\approx}0.44 for
electrons and
\ensuremath{\Delta\beta}$_{h}$\ensuremath{\approx}0.6 for hadrons
and muons. But this contradicts observations (see $\cite{st2}$ and
references there) where the knee in muonic and in hadronic
components is equal to only
\ensuremath{\Delta\beta}$_{h}$\ensuremath{\approx}0.1-0.2.

The problem of primary spectrum recovering from observable EAS
parameters is additionally complicated due to uncertainties in
primaries mass composition. It is very difficult to define primary
particle mass using traditional EAS method. Only hybrid arrays
such as Tibet AS or Chacaltaya array could more or less adequate
solve this problem using emulsion chambers for primary mass
separation. This problem is connected very tightly with the "knee"
problem. If one accepts {\it a priori} a hypothesis of the charge-
or mass-dependent knee in primary spectrum then he accepts also
{\it a priori} the change of mass composition. If experimental
data are processed under this hypothesis then both the knee and
chemical composition change will be ``shown'' by these data after
such processing.

\section{The knee in PeV region: what is responsible for it?}

The question put above could be split in two parts: 1. Is the knee
origin astrophysical or methodical (or any)? 2. Does it caused by
proton or iron primaries? I'll try here to answer the second part
question. As it has been shown in our previous works $\cite{st1}$
there should be observed a break in EAS size spectrum at primary
energy of $\sim$100 TeV/nucleon at sea level. The origin of the
``knee'' is a break of equilibrium (see $\cite{zat}$) between
hadronic and electromagnetic components at a point where the
number of cascading hadrons becomes close to 1. This point is
critical one for EAS development because the number of particle is
discreet value and less than 1 is only 0. Therefore, below this
point the cascade development follows pure electromagnetic
scenario and all EAS parameters change dramatically. Due to spread
of primary masses from A=1 to A=56 there should be observed 2
``knees'': ``proton knee'' at 100 TeV and ``iron knee'' at $\sim$5
PeV. That means the visible knee in PeV region is connected with
iron primary. The knee positions in shower size $N_{e}$ are equal
to $\sim 10^{5}$ and $\sim 10^{6.3}$ consequently. These values
are more or less constant in the current approach and depend
weakly on the altitude. But, the corresponding primary energies
are sure different. Recent results of Tibet AS $\cite{tib}$ on
proton and helium spectra are not understandable in the frames of
commonly used astrophysical knee hypothesis. But, it becomes
absolutely clear in the frames of current approach. Actually, this
hybrid experiment uses different sub-arrays for different
purposes: emulsion chambers are used for core location and primary
mass selection, while the EAS size is measured by the traditional
EAS array. As we noted above, the ``proton knee'' position for
Tibet altitude should be also close to 100 TeV. Therefore, the EAS
size spectrum to the right of this point (their threshold is equal
to 200 TeV) should be steep. To demonstrate this we performed
Monte Carlo simulations with CORSIKA codes.

\section{Results of simulations}

The latest version of CORSIKA program$\cite{hec}$ were used for
calculations(v.6.501). Standard HDPM as well as VENUS and DPMJet
\begin{figure}[htb!]
 \begin{center}
\includegraphics[height=8.cm]{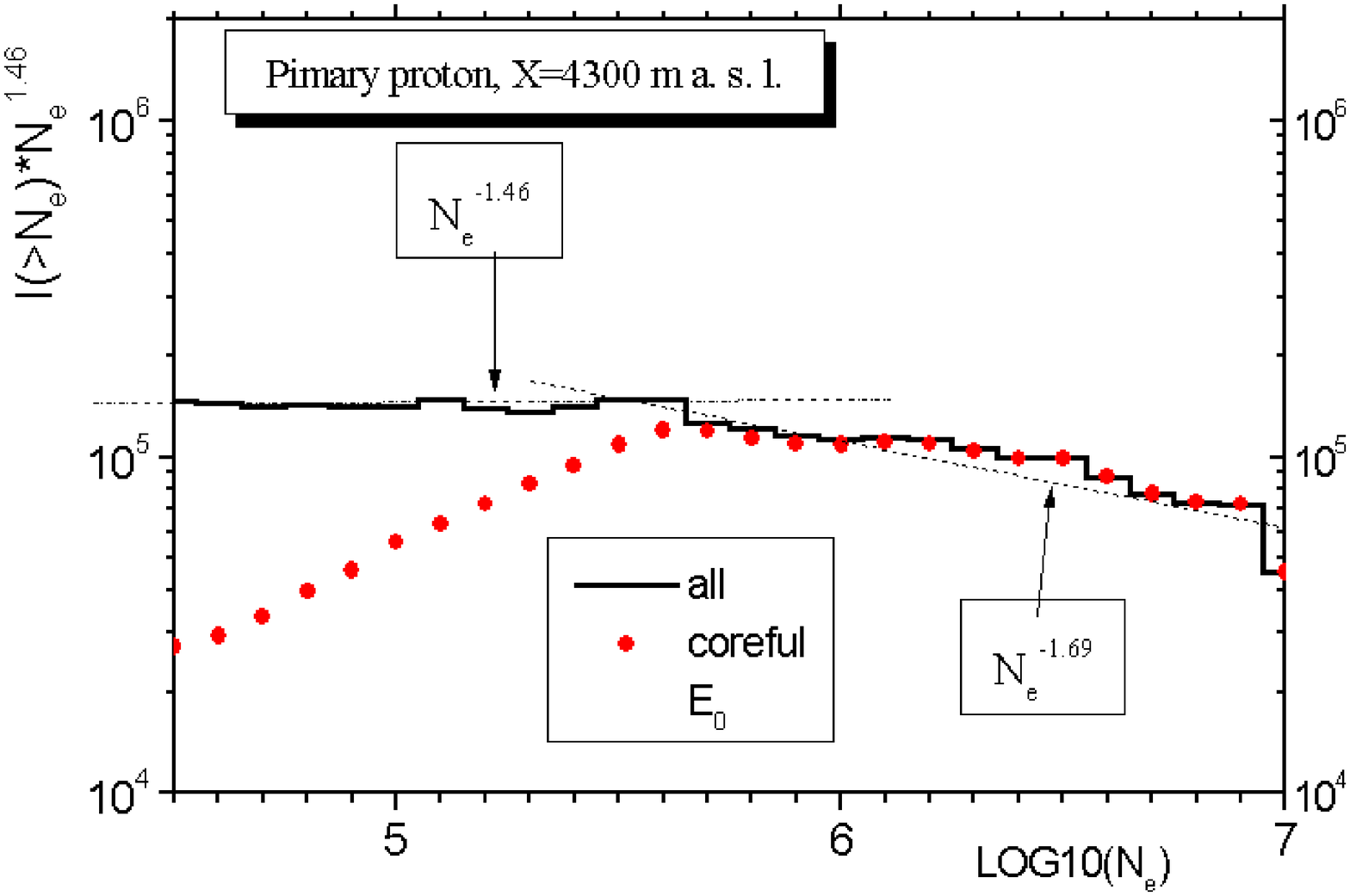}
\includegraphics[height=8.cm]{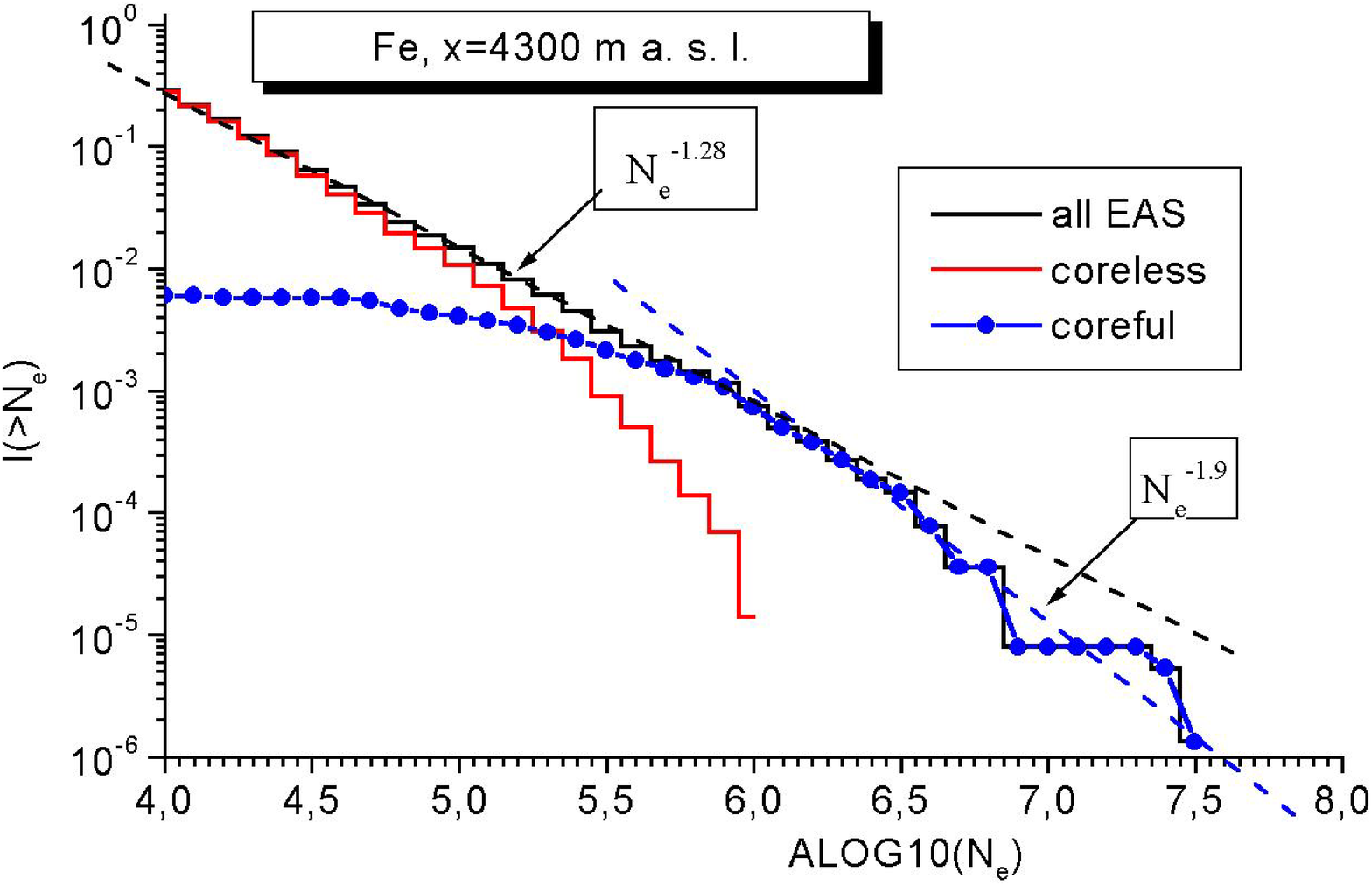}
\end{center}
\caption{Results of Monte Carlo simulations for $N_{e}$
distribution at 4300 m a. s. l. altitude. Top panel for protons
(multiplied by $N_{e}^{1.46}$), low panel for iron nuclei.}
\end{figure}
models were used for high-energy hadron interactions and no
significant difference were seen. Simulations were performed for
proton and iron primary nuclei with the pure power law energy
spectrum having the slope $\gamma= -2.7$ for altitudes from 100 m
to 4.3 Km a. s. l. The number of all electromagnetic particles
($e^{+}, e^{-}$ and $\gamma$) summarized inside radius 1000 m were
assigned to $N_{e}$. Note that such a definition is close but not
equal to the $N_{e}$ usually obtained by experimenters from the
NKG-function. As one can see from Fig.1, the distributions at
altitude 4.3 Km have clear visible kinks at $N_{e}\sim
3\times10^{5}$ for protons and $N_{e}\sim 3\times10^{6}$ for iron.
It is seen that the break of slopes coincides with the appearance
of coreful showers and disappearance of coreless EAS', the curves
for those are also shown. This graph shows that EAS size spectrum
slope becomes steeper at $N_{e}\sim 3.5\times10^{5}$
(corresponding $E_{0}\sim 100 TeV$). Above 200 TeV the slope is
steep enough to explain the Tibet AS data: if one takes
$\alpha=1.2$ then primary spectrum slope
$\gamma=1.69\times1.2=2.03$. This is very close to the value
obtained by Tibet AS. In other words, the spectrum slope measured
by the Tibet AS experiment being recalculated to primary index
taking current approach into account, is very close to $\gamma=
-2.7$ . Therefore, this result is in agreement with direct
spectrum measurements.

A visible kink for iron primaries coincides with the ``knee''
position as one can see in fig. 1 lower panel. In this point our
conclusion is the same as in $\cite{tib}$. Similar shape of the
distributions can be obtained for any other altitudes, but at
different primary energies. The effect depends also on the radius
of integration (on the array dimensions): the smaller radius, the
bigger is effect. Corresponding curves for muons and hadrons have
no visible "knees" $\cite{st1}$.
\begin{figure}[htb!]
\begin{center}
\includegraphics[height=8.5cm]{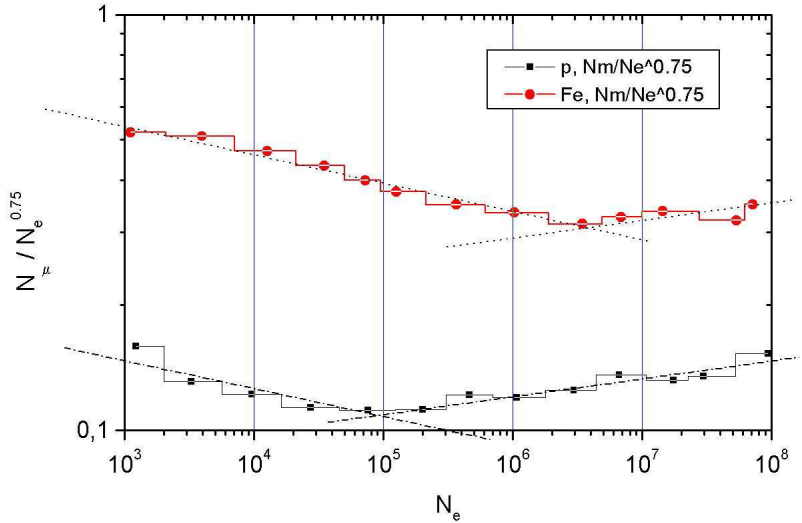}
\end{center}
\caption{Results of Monte Carlo simulations for $N_{e}$ -
$N_{\mu}$ correlation at 1700 m a. s. l. for protons and for iron
nuclei.}
\end{figure}
Due to different behavior of different components as a function of
primary energy, the correlation plot between different EAS
components also exhibits the "knee" as one can see in Fig.2. We
plot these distributions divided by $N_{e}^{0.75}$ to emphasize
the slope change and we made the calculations for another altitude
to show that the effect exists at any altitude. And again the
``knees'' are visible. Only its positions are little bit different
for another altitude. Similar curves obtained experimentally are
usually interpreted as an evidence of the fact that primary mass
composition becomes heavier while here, we obtained it for {\bf
constant mass composition} and for {\bf pure power law spectra}.
This is an example how experimental data could be erroneously
understood and interpreted if one supposes a priori the knee
existence.

\section{Summary}

$\bullet$ EAS size spectrum has "a knee" at any altitude even for
pure power law spectrum of primary cosmic ray.\\
$\bullet$ The index of all-particle primary cosmic ray
energy spectrum does not likely change significantly in a range of 0.1$\div$10 PeV.\\
$\bullet$ The ``knee'' observed experimentally in electromagnetic
EAS component is caused by EAS structure change at energy
\ensuremath{\sim 0.1 PeV / nucleon} where the number of cascading
hadrons becomes equal to zero. Below this energy, EAS's at sea
level are mostly \textit{coreless} while above this
threshold EAS's are mostly \textit{coreful}.\\
$\bullet$ Primary particle mass composition "change" measured by
the EAS method using $N_{\mu}/N_{e}$ ratio is probably methodical
one, while the composition of primaries at the top of atmosphere
could be constant.\\
$\bullet$ The steep spectra of protons and $\alpha$-particles
measured by Tibet AS Group confirm our hypothesis that EAS size
spectra must be steep above the threshold of $\sim$100 TeV /
nucleon, while the primary spectrum does not change a slope.\\
$\bullet$ We coincide with Tibet AS in the conclusion that the
``knee'' in PeV-region is connected with iron nuclei.

\section*{Acknowledgements}

I'd like to thank once again the Developers of CORSIKA program for
very useful instrument for EAS study.

The work was supported in part by RFBR grants Nos 05-02-17395 and
07-02-00964, by the Scientific School Support grant
NSh-4580.2006.2 and by the RAS Basic Research Program "Neutrino
Physics".

\end{document}